\documentclass[prb,a4paper,twocolumn,floatfix,nopacs,nokeys,amsmath,amssymb,nobibnotes,altaffilletter,superscriptaddress]{revtex4}

\usepackage{color}

\usepackage{graphicx}

\begin{document}

\title{Charge density dependent nongeminate recombination in organic bulk heterojunction solar cells}

\author{Daniel Rauh}
\affiliation{ZAE Bayern, Bavarian Center of Applied Energy Research, Am Hubland, 97074 W\"urzburg, Germany}
\author{Carsten Deibel}\email[Electronic mail: ]{deibel@physik.uni-wuerzburg.de}
\affiliation{Experimental Physics VI, Faculty of Physics and Astronomy, Julius-Maximilians-University of W\"urzburg, Am Hubland, 97074 W\"urzburg, Germany}
\author{Vladimir Dyakonov}\email[Electronic mail: ]{dyakonov@physik.uni-wuerzburg.de}
\affiliation{ZAE Bayern, Bavarian Center of Applied Energy Research, Am Hubland, 97074 W\"urzburg, Germany}
\affiliation{Experimental Physics VI, Faculty of Physics and Astronomy, Julius-Maximilians-University of W\"urzburg, Am Hubland, 97074 W\"urzburg, Germany}

\begin{abstract}
Apparent recombination orders exceeding the value of two expected for bimolecular recombination have been reported for organic solar cells in various publications. 
Two prominent explanations are bimolecular losses with a carrier concentration dependent prefactor due to a trapping limited mobility, and protection of trapped charge carriers from recombination by a donor--acceptor phase separation until reemission from these deep states. 
In order to clarify which mechanism is dominant we performed temperature and illumination dependent charge extraction measurements under open circuit as well as short circuit conditions at poly(3-hexylthiophene-2,5-diyl):[6,6]-phenyl-C$_{61}$butyric acid methyl ester (P3HT:PC$_{61}$BM) and PTB7:PC$_{71}$BM (Poly[[4,8-bis[(2-ethylhexyl)oxy]benzo[1,2-b:4,5-b']dithiophene-2,6-diyl][3-fluoro-2-[(2-ethylhexyl)carbonyl]thieno[3,4-b]thiophenediyl]]) solar cells in combination with current--voltage characteristics. We show that the charge carrier density $n$ dependence of the mobility $\mu$ and the recombination prefactor are different for PC$_{61}$BM at temperatures below 300K and PTB7:PC$_{71}$BM at room temperature. Therefore, in addition to $\mu(n)$ a detrapping limited recombination in systems with at least partial donor--acceptor phase separation is required to explain the high recombination orders. 

{\bf This is the pre-peer reviewed version of the following article:\\ D. Rauh, C. Deibel, V. Dyakonov. Adv. Funct. Mater., 2011, 10.1002/adfm.201103118}
\end{abstract}

\maketitle

\section{Introduction}

Organic solar cells (OSC) have already reached efficiencies of 8.3~\% for single junction architectures on lab scale, which is already close to the 10~\% required for module commercialization.\cite{green2011,deibel2010review} Despite significant advances in understanding the fundamental processes in OSC, open questions remain. One of the unresolved issues concerns the exact loss mechanism limiting the device performance: nongeminate recombination of charge carriers. 
Many authors explain their experimental results by Langevin recombination,\cite{langevin1903, deibel2008} in which the annihilation rate of electrons with holes is determined by the low mobility. Nongeminate recombination was also reported to occur via interface states~\cite{cowan2010} or trap states,\cite{kirchartz2011,baumann2011} described by Shockley--Read--Hall (SRH) recombination.\cite{tzabari2011} These processes change the recombination dynamics towards first order processes, as electron and hole concentrations become imbalanced. However, several studies reported recombination orders exceeding two.\cite{montanari2002,shuttle2008, clarke2009, foertig2009}
The recombination rate $R$ can be expressed empirically as
\begin{equation}
	R = k' \bar{n}^\text{order} ,
	\label{eq:rec_rate}
\end{equation}
where $\text{order}$ means the order of decay and $k'$ is the recombination prefactor, here defined to be independent of the charge carrier density $\bar{n}$. In this simple equation, the concentration of electrons $n$ and holes $p$ is not distinguished, as this is usually not possible by experiment. 
For first order recombination, $1/k'$ is the lifetime. For pure Langevin recombination, $k'$ becomes $k=(e/\epsilon)(\mu_n+\mu_p)\approx (e/\epsilon)\mu$, where $e$ is the elementary charge, $\mu$ is the mean mobility of electrons $\mu_n$ and holes $\mu_p$, and $\epsilon$ the effective dielectric constant of the photoactive layer. The physically exact form of the Langevin recombination rate is 
\begin{equation}
	R_L=k (np-n_i^2) \approx k np \approx k \bar{n}^2,
	\label{eq:rec_langevin}
\end{equation}
where $n_i$ is the intrinsic carrier concentration. The last approximation in Eq.~\ref{eq:rec_langevin} assumes $n\approx p$, yielding an order of decay of $2$. 
For apparent recombination orders between one and two, or exceeding two, the prefactor $k_\lambda$ is empirical. The former case can usually be explained by a combination of Langevin and SRH recombination rates. The origin of the higher recombination orders, however, is still under discussion.\cite{clarke2009, foertig2009, shuttle2010}

All approaches to explain the high orders of decay have in common that the recombination process is basically of Langevin type, with the disordered nature of the organic semiconductor blend being accounted for by trapping of charge carriers. The influence of energetic disorder on the charge carrier mobility is already known for decades: the thermally activated hopping process of charge carriers, which can also be described by the multiple-trapping-and-release (MTR) approach, leads to a carrier concentration dependent mobility.\cite{noolandi1977,monroe1985,baranovskii2000} Charge carriers located in the density of states below the transport energy are trapped and immobile (with density $n_t$) whereas charge carriers above (with density $n_c$) are free and have the mobility $\mu_0$. The overall mobility of all charge carriers (with density $n=n_c+n_t)$ corresponds to the measured mobility $\mu$ defined by $\mu_0\cdot n_c=\mu \cdot n$. Nelson~\cite{nelson2003} used one-dimensional Monte Carlo simulations to understand the stretched exponential decays of the charge carrier concentration found in transient absorption experiments. Trapping of charge carriers in the tails of the density of states distribution was found to be responsible for this finding, as it slowed down nongeminate recombination. Recently, Shuttle et al.\cite{shuttle2010}  investigated P3HT:PC$_{61}$BM solar cells experimentally at 300~K. They showed that $k(\bar{n})\propto\mu(\bar{n})$ (Eq.~\ref{eq:rec_langevin}) completely accounted for the carrier concentration of the recombination rate in excess of the expected value of two, i.e., $R=k(\bar{n})\bar{n}^2$ where $k(\bar{n})\propto \bar{n}^\text{(order-2)}$ (cf.~Eq.~\ref{eq:rec_rate}).

In this article, we present experimental evidence that the observed order of decay can only partly be explained by the carrier concentration dependent mobility. We can quantify this discrepancy for P3HT:PC$_{61}$BM solar cells at temperatures below 300~K and for PTB7:$_{71}$BM cells at room temperature. Although the MTR model already includes the existence and influence of trap states we will demonstrate that in addition to the influence on the charge carrier mobility, the donor--acceptor phase separation can protect trapped charge carriers from recombination. The spatial separation of electrons and holes implies that charge carriers trapped within the tail of the density of states distribution cannot be reached by an oppositely charged mobile carrier until the trapped carrier is emitted from the deep state and becomes mobile. Only then can this charge carrier participate in the recombination process. Thus, the emission rates from the trap states slow the recombination rate down even more than the charge carrier mobility, i.e., the impact of trapping on the recombination prefactor alone, can account for. Our model is able to explain the experimentally observed high recombination orders, in contrast to earlier approaches.

\section{Theory} \label{sec:theory}

In order to study recombination processes, it is preferable to analyze the charge carrier density at open circuit conditions for different illumination levels and temperatures. In general, the continuity equation, here shown for electrons, is
\begin{equation}
	\frac{dn}{dt}=-\frac{1}{q}\frac{dj_n}{dx}+G-R ,
\end{equation}
with the time derivative of the electron density $dn/dt$, the spatial derivative of the electron current $dj_n/dx$, the generation rate of free electrons $G$ and the nongeminate recombination rate $R$. In steady state $dn/dt$ is zero. $dj_n/dx$ is approximately zero at open circuit, and cancels with the spatial derivative of the hole current. Consequently, at $V_{oc}$ in steady state all the generated charge carriers  have to recombine ($G=R$). 

The open circuit voltage can be changed by varying the illumination level, i.e., the generation rate. The respective charge carrier density can be measured by different techniques (e.g. photo--CELIV\cite{juska2000}, TPV/TPC\cite{shuttle2008}), here we used a charge extraction (CE) method. Although the CE takes always place under short circuit conditions we distinguish between the charge carrier generation (illumination) under open circuit and short circuit conditions to measure the respective equilibrium charge carrier density. 

The generation rate $G$ can be extracted from the current--voltage (IV) characteristics under illumination with the assumption that at sufficiently high voltages in reverse direction, all generated charge carriers will contribute to the saturated photocurrent density $j_{\text{sat,ph}}$ and will not recombine,
\begin{equation}
G=\frac{j_{\text{sat,ph}}}{qL} .
\label{eq:gen_rate}
\end{equation}
In general the photocurrent density is derived by subtracting the dark IV-curve from the illuminated one, $L$ being the thickness of the photoactive layer. Using $G=R$ and the charge carrier density $\bar{n}$ measured under open circuit conditions, we obtain $R(\bar{n})$ and from the slope $dR/d\bar{n}$ we get the recombination order (Eq.~\ref{eq:rec_rate}). Here we assume that the polaron pair dissociation is independent of the electric field in the device, i.e., that we generate the same number of polarons at $V_{oc}$ and at reverse bias where we calculate $j_{\text{sat,ph}}$. This assumptions is at least justified for P3HT:PC$_{61}$BM between short circuit and open circuit.\cite{limpinsel2010,street2010a,kniepert2011}
 
As we want to find out wether the carrier concentration dependent mobility alone can explain the recombination orders exceeding two, we analyze the different carrier concentration dependent contributions to the Langevin recombination rate (Eq.~\ref{eq:rec_langevin}): we compare the prefactor
\begin{equation}
	k(\bar{n}) = \frac{R_L}{\bar{n}^2} \propto \bar{n}^{\beta} ,
\label{eq:beta}
\end{equation}
determined as described in the previous paragraph, to the independently measured mobility $\mu(\bar{n})\propto \bar{n}^{\alpha}$ as outlined below. If the recombination order can be completely described by the carrier concentration dependent mobility (together with the order of 2 for bimolecular recombination), then $k(\bar{n}) \propto \mu(\bar{n})$, i.e., $\alpha=\beta$ should be found.  For clarity, we drop the bar for the experimentally determined charge carrier concentration from here on and write $n$ instead of $\bar{n}$.

In order to determine the charge carrier density dependence of the mobility, we assume the short circuit current density $j_{sc}$ to be drift-dominated,\cite{shuttle2010}
\begin{equation}
	j_{sc}\approx j_{\text{drift}} = \mu_{\text{drift}}nF .
\label{eq:mu_drift}
\end{equation}
Here $n$ is the charge carrier density measured under short circuit conditions, $F$ the electric field given by the built-in potential. In well optimized solar cells, unbalanced electron and mobilities would inevitably lead to a drop of the $FF$ and the open circuit voltage,\cite{liu2010} which is not the case at least at room temperature. Therefore, we assume  that the mean drift mobility $\mu_{\text{drift}}$ approximately equals the mobility of electrons and holes. We are satisfied with calculating a parameter proportional to $\mu$,
\begin{equation}
	\mu \propto \frac{j_{sc}}{n} \equiv \tilde \mu
	\label{eq:propto_mu}
\end{equation}
for constant $F$, as we are only interested in the carrier concentration dependence, i.e. the exponent $\alpha$,
\begin{equation}
	\mu(n) \propto \tilde \mu(n) \propto n^{\alpha} .
	\label{eq:alpha}
\end{equation}
The same proportionality ($j_{sc}/n$) results assuming the current to be dominated by diffusion, only the proportionality factor is different.

We already pointed out that it is our aim to show that $\alpha$ does not always equal $\beta$, i.e. that the high recombination orders can generally not be solely explained by the carrier concentration dependent mobility. Instead, we will consider a more general recombination rate based on the Langevin rate and prefactor (Eq.~\ref{eq:rec_langevin}),
\begin{eqnarray}
R &=&  \frac{e}{\epsilon }(\mu_n+\mu_p)(n_c+n_t)(p_c+p_t) \nonumber \\
   &=&  \frac{e}{\epsilon } \left( (\mu_n+\mu_p)n_c p_c + \mu_n n_c p_t+ \mu_p p_c n_t + 0 \cdot n_t p_t \right).
\label{eq:cont_trap}
\end{eqnarray}

With the assumptions made before, we simplify this equation to
\begin{eqnarray}
	R  \propto \underbrace{\mu(n)}_{\rm \propto n^{\alpha}} \underbrace{(n_c p_c + n_c p_t+p_c n_t)}_{\rm \propto n^{\beta - \alpha}n^2} \propto n^{\beta+2}.
	\label{eq:cont_trap}
\end{eqnarray}
Here, the charge carrier mobility depends on the carrier concentration due to trapping and release. The first term on the right hand side corresponds to Langevin recombination, but only of mobile carriers. The second and third term are equivalent to SRH recombination, i.e., recombination of a (previously) trapped with a free charge carrier, and is proportional to the mobility of the respective mobile carriers. Therefore, the Langevin prefactor can be adapted for this case as well. As trapped charge carriers are immobile, the term $n_t p_t$ does not contribute to the recombination rate. The advantage of Eq.~\ref{eq:cont_trap} is its higher degree of transparency as compared to the superposition of Langevin and SRH recombination. Bear in mind that free and trapped charge carriers are not completely separate reservoirs of charge carriers, but that free carriers can be trapped and trapped ones can be reemitted as described by the MTR framework. Due to Fermi--Dirac statistics, $n_t>n_c$ in steady state. We already point out here that in the case of a significant donor--acceptor phase separation, recombination can only take place at the heterointerface. Thus, only mobile charge carriers could recombine directly, as trapped charge carriers would be protected from recombination within their respective material phase. Thus, Eq.~\ref{eq:cont_trap} would simplify to $R \propto \mu(n) n_c p_c$, where the free carrier reservoirs are refilled by reemission from trapped charge carriers, this thermal activation process becoming the limiting factor for the recombination rate instead of the mobility. This case leads to a slower decay of the overall charge carrier concentration, corresponding to a higher recombination order going beyond the impact of the mobility.

\section{Results and Discussion}
\subsection{Results}
\begin{figure}[tb] 
   \centering
   \includegraphics[scale=0.5]{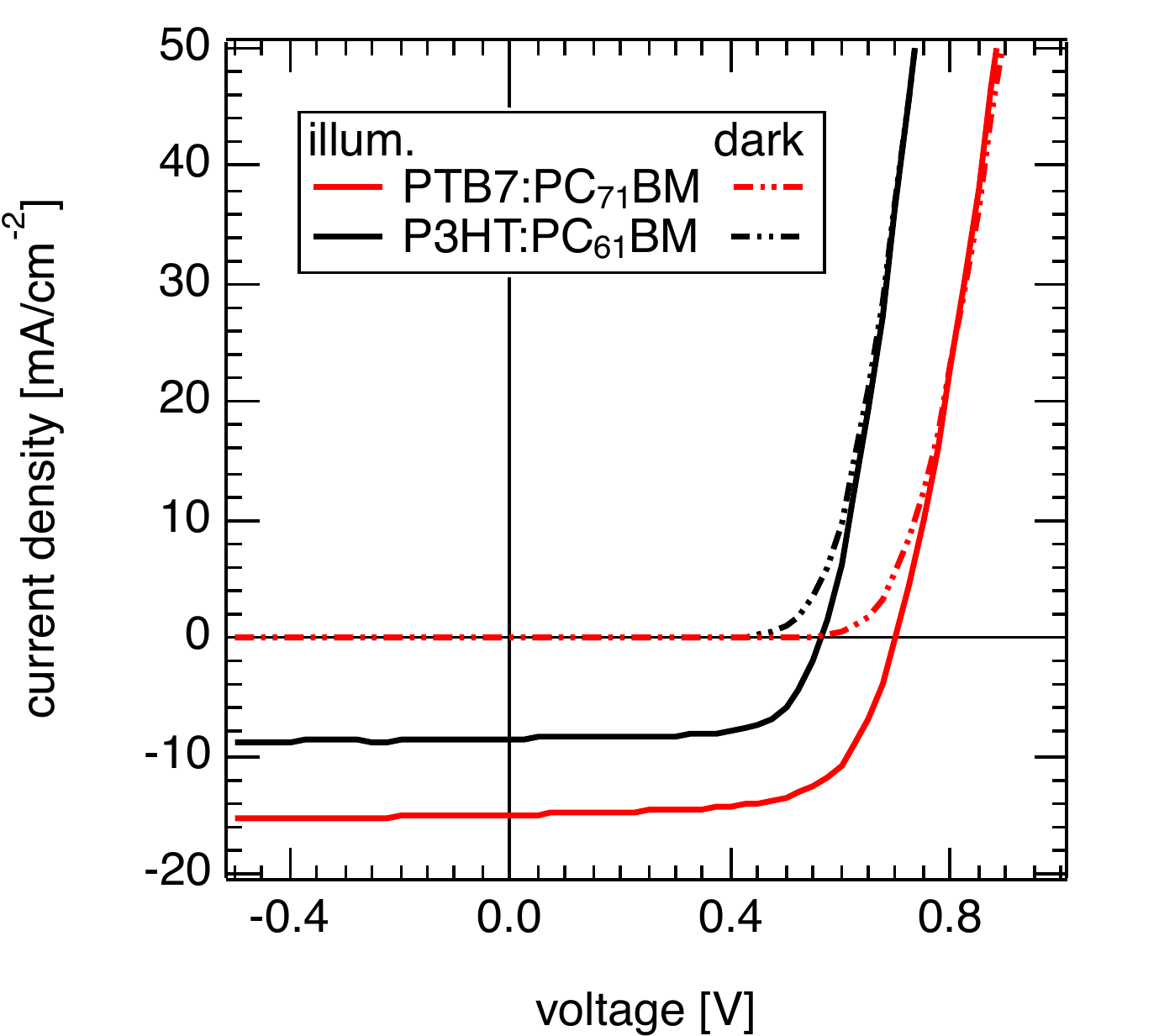} 
          \caption{Current--voltage characteristics of P3HT:PC$_{61}$BM and PTB7:PC$_{71}$BM solar cells under 1 sun illumination conditions and in the dark.}
   \label{fig:iv_1sun}
\end{figure}

All CE and IV measurements were performed on P3HT:PC$_{61}$BM and PTB7:PC$_{71}$BM solar cells showing photovoltaic behavior comparable with literature.\cite{brabec2010,liang2010} The P3HT:PC$_{61}$BM cell had an efficiency of $\eta$=3.4~\% ($V_{oc}$=570~mV, $J_{sc}$=8.6~mA/cm$^2$, $FF$=69~\%) under 1~sun illumination, the PTB7:PC$_{71}$BM $\eta$=7.0~\% ($V_{oc}$=700~mV, $J_{sc}$=15.0~mA/cm$^2$, $FF$=67~\%). The IV curves in dark and for 1 sun illumination are shown in Fig.~\ref{fig:iv_1sun}. 
Details about the calibration of the solar simulator as well as the performed CE measurements are given in the Experimental Section.

\begin{figure}[tb] 
   \centering
   \includegraphics[scale=0.5]{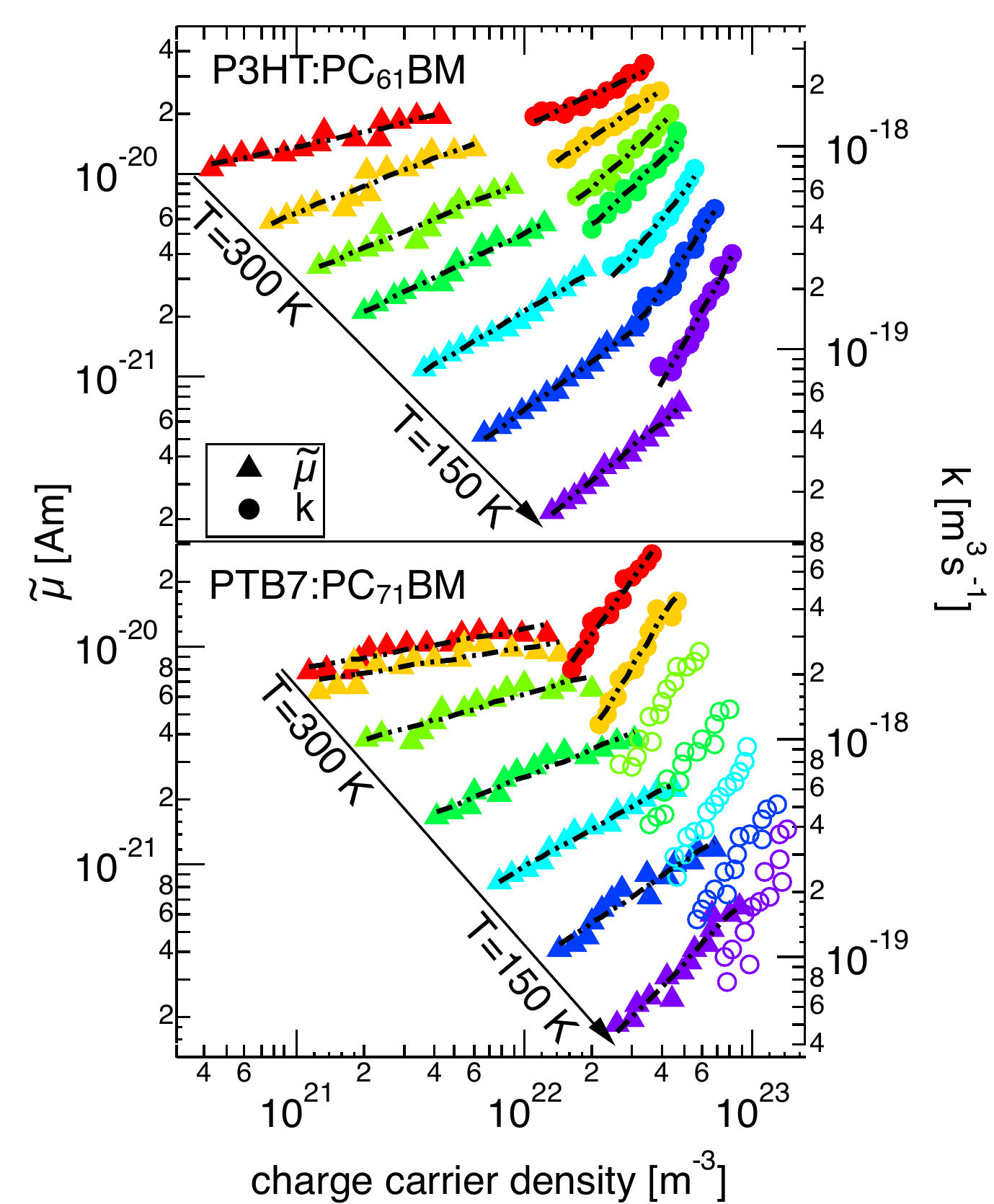} 
          \caption{Charge carrier dependence of $\tilde \mu$ (Eq.~\ref{eq:propto_mu}) (triangles, left axis) and the Langevin recombination prefactor k (circles, right axis) for P3HT:PC$_{61}$BM (top) and PTB7:PC$_{71}$BM (bottom). Solid circles were used for temperatures where the contacts are ohmic, open circles where the cell was limited by injection barriers. Details are given in the text. The dashed--dotted lines indicate the fits to Eq.~\ref{eq:alpha} and Eq.~\ref{eq:beta}., respectively.}
   \label{fig:kbr_mu}
\end{figure}

Fig.~\ref{fig:kbr_mu} shows $\tilde\mu(n)$ and $k(n)$ obtained from CE and IV measurements for light intensities ranging from 0.01 to 0.5~suns. 
For both material systems and at all temperatures the power law dependence of $k(n)$ and $\tilde \mu(n)$ as described by Eq.~\ref{eq:beta} and Eq.~\ref{eq:alpha} can be observed. 
The slopes of $\tilde \mu(n)$ and $k(n)$ are different, implying that the charge carrier density dependence of the mobility alone cannot explain the $k(n)$ dependence.

The recombination prefactor $k(n)$ was calculated using  Eq.~\ref{eq:rec_langevin} and Eq.~\ref{eq:gen_rate}, where for $j_{\text{sat,ph}}$ we used the value at -3.5~$V$. At this negative bias the photocurrent was saturated for all temperatures which was proven by the linear dependence of the calculated generation rate vs. the light intensity $P_L$. The exponent $\delta$ from $G \propto P_L^{\delta}$ ranged from 0.95 at 300~K to 0.92 at 150~K for P3HT:PC$_{61}$BM and from 0.95 to 0.94 for PTB7:PC$_{71}$BM. Recombination losses during the extraction in the CE measurements in the range of only a few percent and therefore not taken into account. The steady state charge carrier density under short circuit conditions was always lower than under open circuit conditions at the same illumination level because of the continous sweep out of charge carriers before extraction. 

\begin{figure}[tb] 
   \centering
   \includegraphics[scale=0.5]{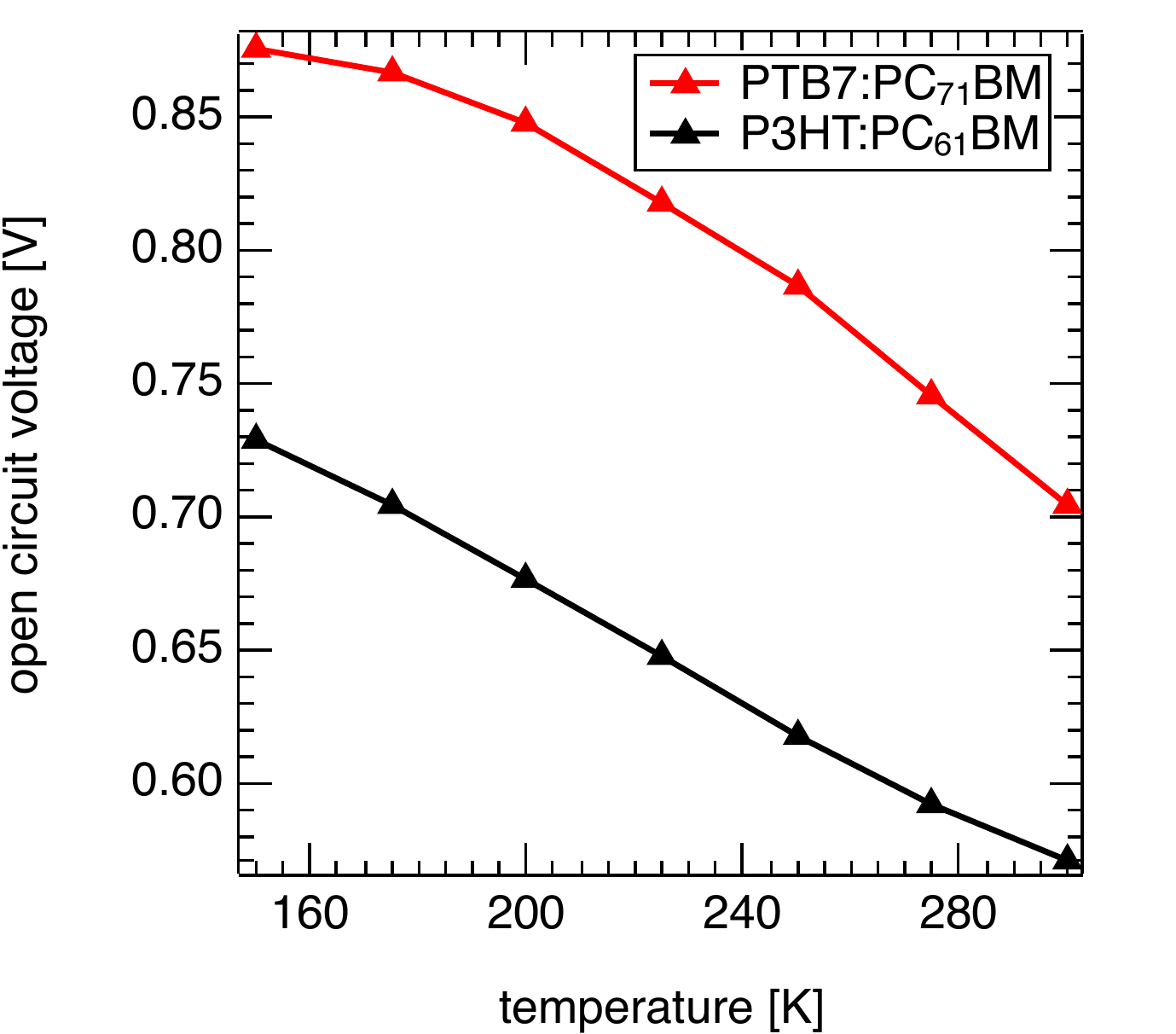} 
   \caption{The open circuit voltage versus temperature at 1~sun. For the P3HT:PC$_{61}$BM solar cell the contacts are not limiting the $V_{oc}$ justified by the linear behavior dependence, for the PTB7:PC$_{71}$BM solar cell, the contacts influence the $V_{oc}$, especially at lower temperatures which becomes visible by a saturation of $V_{oc}$.}.
   \label{fig:voc_T}
\end{figure}

The $V_{oc}(T)$ dependence for both material systems is shown in Fig.~\ref{fig:voc_T}, where the linear increase of the open circuit voltage with decreasing temperature in the case of P3HT:PC$_{61}$BM indicates that the contacts have an ohmic behavior with negligible injection barriers.\cite{rauh2011} Therefore, $V_{oc}$ is directly related to the charge carrier density at open circuit conditions by
\begin{equation}
V_{oc}=\frac{E_g}{q}+\frac{n_{id} k_B T}{q}\ln\left(\frac{np}{N_c^2}\right)\quad{,}
\label{eq:vocT}
\end{equation}
with $E_g$ the effective band gap, $n_{id}$ the ideality factor, $k_B$ Boltzmann's constant and $N_c$ the effective density of states.\cite{cheyns2008,koster2005} The situation changes regarding the PTB7:PC$_{71}$BM solar cell, where we observe a linear increase of $V_{oc}$ with decreasing temperature only in the range of around 250 to 300~K, whereas at lower temperatures the open circuit voltage is limited by injection barriers that are dominant then.\cite{rauh2011} In this regime Eq.~\ref{eq:vocT} is not valid and the measured charge carrier density can also be affected by the barriers. Hence, we  did not further evaluate the $k(n)$ data in the temperature range of 150 to 250~K for PTB7:PC$_{71}$BM, as indicated in Fig.~\ref{fig:kbr_mu} by open instead of solid circles.

\begin{figure}[tb] 
   \centering
   \includegraphics[scale=0.5]{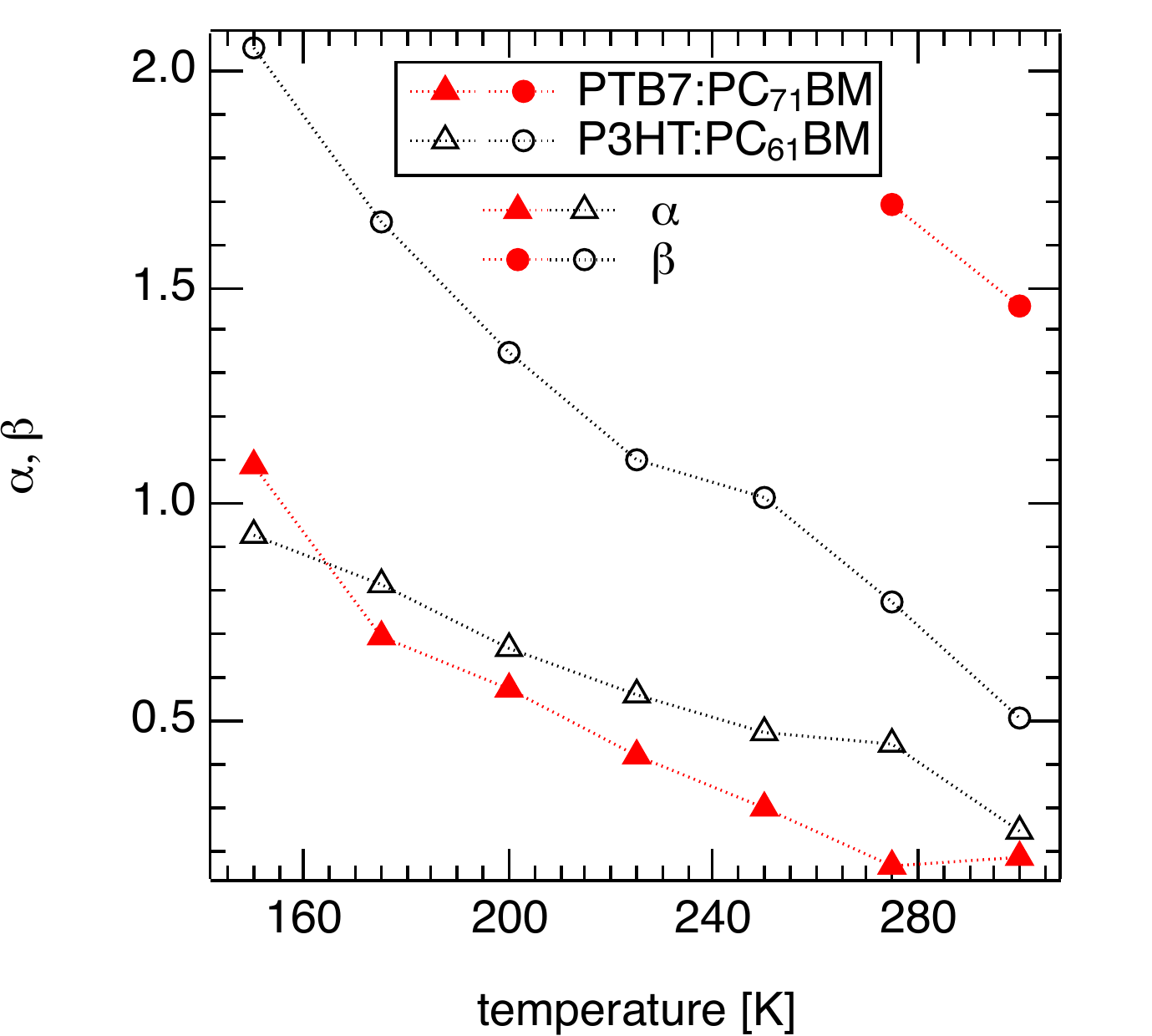} 
          \caption{The values of $\alpha$ and $\beta$ obtained from Fig.~\ref{fig:kbr_mu} using Eq.~\ref{eq:alpha} (triangles) and Eq.~\ref{eq:beta} (circles) for the P3HT:PC$_{61}$BM (black) and PTB7:PC$_{71}$BM (red) versus temperature. $\beta$ for PTB7:PC$_{71}$BM at lower temperatures was not evaluated because of the contact limitation of $V_{oc}$ in this temperature range.}
   \label{fig:lambda}
\end{figure}

$\beta$ was obtained from the data presented in Fig.~\ref{fig:kbr_mu} by using the slope of the linear fit of ln($k$)~vs.~ln($n$) over the whole data set (Eq.~\ref{eq:beta}), $\alpha$ from the $\ln(\tilde \mu)$ vs.\ $\ln(n)$ fit (Eq.~\ref{eq:alpha}). The fits are presented in Fig.~\ref{fig:kbr_mu} as dash--dotted lines. The calculated values of $\beta$ and $\alpha$ are summarized in Fig.~\ref{fig:lambda}. The case of $\beta = \alpha$ would imply that the recombination orders higher than two can be explained completely by pure Langevin recombination with an additional charge carrier density dependent mobility. 

\subsection{Discussion}

For the P3HT:PC$_{61}$BM solar cell at 300K (Fig.~\ref{fig:kbr_mu}), the mobility ($\alpha$) shows nearly the same carrier concentration dependence as the recombination prefactor $k$ ($\beta$), i.e.\ the apparent recombination order above two is due to the charge carrier concentration dependent mobility in accordance with literature.\cite{shuttle2010} At lower temperatures, however, it becomes clear that $\alpha$ and $\beta$ are not increasing by the same amount with lower temperatures, resulting in $\alpha < \beta$. The temperature behavior of $\beta$ for P3HT:PC$_{61}$BM (Fig~\ref{fig:lambda}), determined independently, is in accordance with literature.\cite{foertig2009}

The 275 and 300~K data of PTB7:PC$_{71}$BM hold a clear statement. Whereas the values of $\alpha$ are similar to those of the P3HT:PC$_{61}$BM device, $\beta$ is much higher. This results in a high discrepancy between $\beta$ and $\alpha$ even at room temperature for PTB7:PC$_{71}$BM in contrast to P3HT:PC$_{61}$BM .

In both material systems the charge carrier density dependence of the mobility alone cannot explain the recombination order being higher than two and a trap-assisted recombination mechanism has to be taken into account.

Generally, the temperature dependence of the carrier concentration dependent mobility (i.e., $\alpha$) is experimentally not well investigated. Tanase et al.\cite{tanase2003} showed charge carrier dependent mobility data of P3HT diodes and field effect transistors (FET). In the regime of low charge carrier densities ($1\cdot 10^{21}-4\cdot 10^{22}$~m$^{-3}$) occurring in solar cells under standard light intensities, they observed almost no $\mu(n)$ dependence by determining the mobility in the space charge limited regime. The data from the FET measurements at higher charge carrier densities ranging from $2\cdot10^{24}-3.5\cdot10^{25}$~m$^{-3}$ showed a clear $\mu(n)$ dependence. Overall they proposed $\mu \propto n^{T_{0}/T-1}$, with $T_0$ the width of an exponential density of states. This implies an increasing $\alpha$ with decreasing temperature, which is in good agreement with our measurements. The same trend for $\alpha(T)$ was predicted by Pasveer et al.\ from numerical simulations of the hopping transport in a master equation approach.\cite{pasveer2005}

The temperature dependence of the difference between $\mu(n)$ (i.e., $\alpha$) and $k(n)$ (i.e., $\beta$) reinforces our view\cite{baumann2011} that at lower temperatures the influence of a trapping on recombination becomes more pronounced. Under these conditions the release of a trapped charge carrier into a transport state is less probable than at room temperature, as the emission is thermally activated by a Boltzmann factor. The existence of a broad distribution of traps in P3HT:PC$_{61}$BM ranging from 20 to 400~meV was confirmed by thermally stimulated current (TSC) technique with the distribution maximum at 105~meV.\cite{schafferhans2010}

Considering the contributions to the recombination rate discussed in section~\ref{sec:theory} in the context of Eq.~\ref{eq:cont_trap}, we can separate three contributions, two of which are directly apparent. 
\begin{enumerate}
\item The recombination of mobile charge carriers $n_c p_c$  corresponds to the classical Langevin picture, with the difference that not all charge carriers participate. Depending on the dynamics of trapping and emission, the order of recombination in view of the \emph{overall} carrier concentration $n$ can exceed the value of two. An additional contribution to the carrier concentration dependence is due to the recombination prefactor, i.e., the mobility $\mu(n)$, as described e.g. by Nelson.\cite{nelson2003}
\item The recombination of mobile charge carriers with trapped ones, $n_c p_t$ and $p_c n_t$. The prefactor is proportional to the mobility of the free carriers,\cite{kirchartz2011} leading again to a recombination rate in accordance with Langevin theory. Alternatively, this contribution can be described by SRH, although the carrier concentration dependence of the mobility does not automatically follow from the trap population---in contrast to MTR. Accordingly, more fit parameters are required for SRH.
\item Contributions to recombination of free with free and free with trapped charge carriers, as described in the prior two points, considering (partial) phase separation. Experimentally, phase separation has been reported at least for P3HT:PCBM, see e.g. Ref.~\onlinecite{agostinelli2011}. Charge carriers trapped within their respective material phase cannot be reached by their oppositely charged recombination partners residing in the other material. Only upon thermal activation of the trapped charge carriers from the deep states are they able to recombine at the heterointerface. This third contribution therefore increases the first one, $n_c p_c$, combined with a reduced second contribution due to the phase separation. Consequently, the recombination rate is decreased due to the slow emission process, leading to high orders of decay.
\end{enumerate} 

We point out that for P3HT:PC$_{61}$BM  the impact of trapping is more pronounced for lower temperatures due to a higher effective disorder, i.e. a much slower emission rate from traps, directly influencing the third contribution. Similarly, in P3HT:bisPC$_{61}$BM films a significantly slower decay of the polaron signal than for P3HT:PC$_{61}$BM was found by transient absorption measurements.\cite{faist2011} This corresponds to a higher apparent recombination order and is consistent with bisPC$_{61}$BM exhibiting more and deeper traps than PC$_{61}$BM, as observed by TSC.\cite{schafferhans2011}

Clearly, trap states have a strong impact on the apparent recombination order, in terms of the charge carrier mobility and the populations of charge carriers which are available for recombination. The impact on the device performance needs to be considered accordingly.

\section{Conclusions}
We determined the charge carrier dependence of the mobility $\mu$ and the recombination prefactor $k$ for various temperatures by independent experimental techniques. For P3HT:PC$_{61}$BM solar cells at 300~K, the mobility showed nearly the same dependence as $k$ on the charge carrier density in accordance with literature.\cite{shuttle2010} At lower temperatures the discrepancy between the $\mu(n)$ and $k(n)$ dependency increased. Investigations of a highly performing PTB7:PC$_{71}$BM solar cell showed the discrepancy between $\mu(n)$ and $k(n)$ dependency already at room temperature.  Our findings substantiate the proposition that not only the impact of trapping on the recombination prefactor, proportional to the charge carrier mobility, is responsible for increasing the order of charge carrier decay beyond the value of two expected for bimolecular recombination. Instead, for systems with (partial) phase separation, trapped charge carriers can be protected from recombination. Only after their thermally activated release from the deep states are they able to contribute to the recombination rate, leading to an additional increase of the recombination order. This scenario implies that in PTB7:PC$_{71}$BM the influence of trapping in combination with phase separation is more significant than in P3HT:PC$_{61}$BM, lowering the charge carrier recombination rate. We expect apparent recombination orders greater than two to be an inherent property of disordered organic semiconductor blends.

\section{Experimental Section}
\small{The cells were processed as follows. Structured indium tin oxide (ITO)/glass was cleaned successively in soap water, acetone and isopropanol for at least 10~min in an ultrasonic bath before a thin layer of poly(3,4-ethylendioxythiophene):polystyrolsulfonate (PEDOT:PSS, CLEVIOS P VP AI 4083) was spincoated to serve as anode. After transferring the samples into a nitrogen filled glovebox  a heating step of 130 $^\circ C$ for 10~min was applied. The active layers were spincoated from chlorobenzene solutions using blend ratios of P3HT:PC$_{61}$BM = 1:0.8 and PTB7:PC$_{71}$BM = 1:1.5, resulting in film thicknesses of 200~$nm$ (P3HT:PC$_{61}$BM) and 105~$nm$  (PTB7:PC$_{71}$BM). The P3HT:PC$_{61}$BM film was annealed again for 10~min at 130~$^\circ C$, the PTB7:PC$_{71}$BM was left as cast.  Afterwards the metal contacts Ca(3~$nm$)/Al(120~$nm$) were thermally evaporated at a pressure below $1\cdot 10^{-6}$~mbar. PC$_{61}$BM and PC$_{71}$BM was purchased from Solenne, P3HT (P200) from Rieke Metals and PTB7 from 1-material.
 
Solar cell efficiencies were measured directly after the metal contact evaporation in a nitrogen filled glovebox with a Keithley 237 SMU. For illumination an Oriel 81160 AM1.5G solar simulator was used calibrated to 100~mW/cm$^2$ by a filtered silicon reference cell (PV measurements, Inc ). To calculate the mismatch factor $M$ we used a homemade system to measure the external quantum efficiency EQE of the test cells. We note that the EQE was measured without backlight illumination and not at the same cells as the ones shown in the article but with similar thicknesses and photovoltaic behavior. The spectra of the solar simulator was measured using a calibrated spectrometer (GetSpec 2048).  Since the mismatch was close to one for both material systems (M=0.96 for P3HT:PC$_{61}$BM, M=1.04 for PTB7:PC$_{71}$) we did not readjust the solar simulator for an exact efficiency determination. This was done because the error of spatial inhomogeneity of the light beam was in the same range. 

All temperature dependent current--voltage (IV) characteristics and charge extraction 
measurements  were performed in a closed cycle cryostat (Janis CCS 550) with He as contact gas. For measurements under illumination a 10~W high power white light emitting diode (Seoul) was used. The light intensity was matched to the short circuit current obtained from the measurements with the solar simulator and defined than to have 1~sun. The light was varied by changing the current driving the LED as well as a filter wheel using a set of neutral density filters.

IV-measurements were recorded by a SMU (Keithley 2602) to extract $V_{oc}$. $n$ was determined by the CE technique, which is described in detail elsewhere.\cite{shuttle2008b} The used CE setup consisted of a function generator (Agilent 81150A) for applying $V_{oc}$ in case of the measurement of $n_{oc}$ to the solar cell and triggering a transistor for switching the LED on and off. The CE signal was preamplified by a current amplifier (FEMTO DHPCA-100) before it was detected with an oscilloscope (Agilent DSO 90254A). Integrating the obtained signal over time resulted in the extracted charges at $V_{oc}$. This value was then corrected by the number of charges stored at the electrodes when applying the open circuit voltage. The capacitance was measured by charge extraction experiments in the dark in reverse bias. To calculate the density of extracted charge carriers the extracted charge was divided by the volume of the solar cell bulk and the elementary charge.}

\section*{Acknowledgements}
The current work is supported by the Deutsche Forschungsgemeinschaft in the framework of the PHORCE project (Contract No.~DE 830/8-1). D.R.'s work was financed by the Dephotex Project within the 7th Framework Programme of the European Commission. C.D.\ gratefully acknowledges the support of the Bavarian Academy of Sciences and Humanities. V.D.?s work at the ZAE Bayern is financed by the Bavarian Ministry of Economic Affairs, Infrastructure, Transport and Technology.

\end{document}